\RequirePackage{fix-cm}
\documentclass[twocolumn,epjc3]{svjour3}  
\smartqed  
\RequirePackage{graphicx}
\journalname{Eur. Phys. J. A}
\begin{document}

\title{Constructing probability density function of net-proton multiplicity distributions using Pearson curve method
}


\author{Nirbhay Kumar Behera\thanksref{e1,addr1}
        \and
        Min Jung Kweon\thanksref{addr2} 
}

\thankstext{e1}{e-mail: nirbhaykumar@cutn.ac.in}


\institute{Department of Physics, Schools of Basic and Applied Sciences, Central University of Tamil Nadu, Thiruvarur, India-610005 \label{addr1}
           \and
           Inha University, 100, Inharo, Nam-gu, Incheon, South Korea-22212 \label{addr2}
          }

\date{Received: date / Accepted: date}

\maketitle

\begin{abstract}
The probability density functions of proton, anti-proton, and net-proton multiplicity distributions are constructed from the Beam Energy Scan results of the STAR experiment using the Pearson curve method. The constructed distributions of proton and anti-proton are compared with Poisson and Binomial distributions. The net-proton probability distributions are compared with Skellam distributions to study the O(4) criticality near the chiral crossover transition. The $C_{6}/C_{2}$ results estimated from the obtained PDFs are compared with Skellam and Binomial baselines for the Beam Energy Scan data. The current study shows some signatures of O(4) criticality, which can be further investigated by precision measurements of the cumulants and understanding the contribution of non-critical fluctuations to them. This study also provides a baseline for the higher order cumulant measurement in the upcoming RHIC BES II program and future LHC run.
\end{abstract}
%
%
\section{Introduction}
\label{intro}
One of the most intriguing subjects in the theory of Quantum Chromodynamics (QCD) is mapping its phase diagram. In this context, we have only a schematic picture of the QCD phase diagram. It was first
predicted that at the limit of vanishing quark masses and baryochemical potential ($\mu_B = 0$), the transition is likely to be second order, belonging to the ${\it O}$(4) universality class of 3-dimensional symmetric spin model \cite{Pisarski:1983ms}. Current lattice QCD calculation has shown that at vanishing $\mu_B$, the chiral and deconfinement transition is a smooth crossover on the temperature ($T$) axis \cite{Aoki:2006we}. But at non-vanishing $\mu_B$, the phase transition will be of first order \cite{Ejiri:2008xt,Bowman:2008kc,Stephanov:2007fk}. Thus, various effective field theory models suggest the presence of a critical point (CP) where the second order phase transition line ends and first order phase transition line starts \cite{Asakawa:1989bq,Hatta:2002sj,Scavenius:2000qd,Halasz:1998qr}. Hence, locating this QCD critical point in the $T-\mu_B$ plane has drawn much attention in recent times. Due to the fermion sign problem at finite $\mu_B$, lattice QCD faces tremendous challenges to establishing the fact of the presence or absence of such CP. 
\par
Meanwhile, various theoretical works have proposed that if indeed there is a CP in the $T-\mu_B$ plane, it can be observed experimentally by varying the collision energy \cite{Stephanov:1999zu}. Event-by-event measurement of fluctuations of conserved charges multiplicities is an excellent tool for such study \cite{Stephanov:1998dy,Stephanov:2008qz}. Then any nonmonotonic behavior of higher-order cumulants and their ratios as a function of $\sqrt{s_{NN}}$ is proposed as the signature of the presence of CP \cite{Stephanov:1999zu,Stephanov:2008qz}. Moreover, measurement of higher-order cumulants are potential probes for pseudo-critical behavior near the vicinity of chiral crossover phase transition and determining the freeze-out parameters \cite{Karsch:2010ck,Friman:2011pf,Bazavov:2012vg,Borsanyi:2013hza,Borsanyi:2014ewa}. Various experiments, like Beam Energy Scan (BES) program at Relativistic Heavy-ion collision (RHIC) at BNL, the Super Proton Synchrotron (SPS) at CERN and Compressed Baryonic Matter (CBM) experiment at FAIR, GSI aim for such study.
\par
Net-proton multiplicity fluctuations are regarded as a proxy to the net-baryon number fluctuations. The coupling strength of proton with the sigma field is more than the charged particle (pions) \cite{Stephanov:2008qz,Hatta:2003wn}. Hence, measurement of higher-order cumulants of net-proton multiplicity fluctuations is an excellent tool to search for critical phenomena. Beam Energy Scan (BES) results of cumulants ($C_{n}$) of net-proton multiplicity distributions up to fourth orders are reported by STAR experiments at RHIC \cite{Adamczyk:2013dal,Luo:2015ewa,STAR:2020tga,STAR:2021iop}. Here $C_{n}$ represents the $n^{th}$ order cumulant. The most recent energy dependence results of $C_{4}/C_{2}$ shows a nonmonotonic behaviour.  In the energy range $\sqrt{s_{NN}}= 7.7 - 62.4$ GeV, the most central collision data show a nonmonotonic variation of a significance of 3.1$\sigma$ with the collision energy \cite{STAR:2020tga,STAR:2021iop}. In contrast, a hadronic-transport model (UrQMD\cite{Bleicher:1999xi}) and different variants of hadron resonance gas (HRG) models, which do not include CP show monotonic energy dependence. Additionally, large values of four-particle correlation function of protons are obtained for $\sqrt{s_{NN}} <$  19.6 GeV, which could be a signature of a CP or first order phase transition \cite{Bzdak:2016jxo}. However, data for $\sqrt{s_{NN}} \leq$ 27 GeV are associated with relatively large uncertainties, which will be reinvestigated by the upcoming BES-II program.
\par
Meanwhile, it is demonstrated that the criticality belonging to the O(4) and Z(2) universality class in 3-dimensional symmetric spin model at zero and non-zero $\mu_{B}$ can be studied by knowing the probability distribution of net-baryon (net-proton) number \cite{Morita:2013tu,Morita:2012kt,Morita:2014fda}. It is shown that the characteristic shape of the probability distribution is influenced by the nature of phase transition, and it further reveals the universality class of the transition. For example, the probability distribution appears narrower than the Skellam near the chiral crossover transition belonging to O(4) universality class \cite{Morita:2014fda}. The narrowing of the probability distributions also leads to a negative value of $C_{6}$. So the O(4) criticality will be reflected both in the probability distributions and in the sixth order cumulants. In Ref. \cite {Morita:2014fda}, the O(4) criticality is investigated using the efficiency uncorrected net-proton distributions measured by the STAR experiment, which shows some qualitatively similar features as the model. This work further stresses to carry out a similar study with the efficiency corrected data. However, the efficiency corrected net-proton distribution is not obtained by the experiment. 
\par
To address this demand, for the first time, the Pearson curve method (PCM) is employed for constructing the probability density function (PDF) of proton, anti-proton and net-proton multiplicity distributions from the efficiency corrected first four cumulants (moments) data. The methodology of the approximation of PDF from the first four cumulants is discussed in Section II. In Section III, this technique is used to obtain the PDF of proton, anti-proton, and net-proton multiplicity distributions for the BES data. Then from the obtained probability density functions, the ratio with respect to Skellam distribution is obtained to test the O(4) criticality. Then from the constructed PDFs the $C_6$ and its ratios to $C_{2}$ are estimated as a function of $\sqrt{s_{NN}}$ for most central collision data. The results are also compared with Skellam and Negative Binomial distribution expectations. The summary and outlook are discussed in Section IV.

\section{Pearson curve method}
\label{sec:1}

A statistical distribution or frequency data is characterized by mean ($\mu$),
sigma ($\sigma$), skewness ($S$), kurtosis ($\kappa$) or alternatively
by first four moments (cumulants). In 1895, English mathematician Karl
Pearson devised a method for deriving the functional form of a PDF
from the first four moments of a given frequency data \cite{Pearson343,Pearson443}. He proposed a family of distributions, also known as a system of curves, which
includes famous distributions, like Normal distribution, Beta,
Gamma and $\chi^2$ distributions. According to Pearson, the PDF
$f(x)$, of any of this system of curves can be derived, if it satisfies
the following differential equation \cite{Pearson343,Pearson443,statbook}. 
\begin{equation}
\frac{1}{f(x)} \frac{df(x)}{dx} = - \frac{a + x} { b_{0} +
  b_{1}x + b_{2} x^{2}},
\label{diffEq}
\end{equation}
where $a, b_{0}, b_{1}, b_{2}$ are constant parameters of
the distribution. These constant parameters are specific functions of the first
four moments of a distribution. Generally, the parameters are
expressed by central moments as follows \cite{statbook,pearsontypes}. 
\begin{eqnarray}
\label{param1}
a &=& \frac{\sqrt{m_{2}}\sqrt{\beta_{1}}(\beta_{2} + 3)}{10\beta_{2} - 12\beta_{1} -
      18}\\
\label{param2}
b_{0} &=& \frac{m_{2}(4\beta_{2} - 3\beta_{1})}{10\beta_{2} - 12\beta_{1} -
      18}\\
\label{param3}
 b_{1} &=& a\\
\label{param4}
b_{2} &=& \frac{2\beta_{2} - 3\beta_{1} - 6}{10\beta_{2} - 12\beta_{1} -
      18},
\end{eqnarray}
where $\beta_{1} = m_{3}^{2}/m_{2}^{3}$ = $S^2$, $\beta_{2} =
m_{4}/m_{2}^{2}$ = $\kappa +3$, and $m_{2}, m_{3}, m_{4}$ are second, third and
fourth central moments, respectively. Eq. (\ref{diffEq}) yields the PDF with mean value zero. For the PDF with non$-$zero mean, Eq. (\ref{diffEq}) is modified as follows \cite{mathematica}.
\begin{equation}
\frac{1}{f(x)} \frac{df(x)}{dx} = - \frac{a_{0} + a_{1} x} { b_{0} +
  b_{1} x + b_{2} x^{2}}~,
\label{diffEqN}
\end{equation}
where the modified constant parameters are given as follows.
\begin{eqnarray}
\label{param5}
a_{0} &=& 2\mu(5\beta_2 - 9 ) - \sigma \sqrt{\beta_1} (3+\beta_2) \nonumber\\
&& -12\mu\beta_{1} \\
\label{param6}
a_{1} &=& 18 + 12\beta_{1} - 10\beta_{2}\\
\label{param7}
b_{0} &=& 3(\mu^{2} + \sigma^{2})\beta_{1} - 2(\mu^{2} + 2\sigma^{2})\beta_{2} \nonumber\\ && + 6\mu^{2} + \mu\sigma\sqrt{\beta_1}(3+\beta_2)\\
\label{param8}
 b_{1} &=& 4\mu(\beta_{2} - 3) -6\mu\beta_{1} - \sigma\sqrt{\beta_1}(3+\beta_2)\\
\label{param9}
b_{2} &=& 6 + 3\beta_{1} - 2\beta_{2}~.
\end{eqnarray}

The PDF obtained using Eq.(\ref{diffEqN}) has the same form as obtained from Eq.(\ref{diffEq}) except the distribution is shifted by the non$-$zero value of $\mu$. Due to the shift-invariance property of central moments and cumulants, PDF obtained from Eq.(\ref{diffEq}) and Eq.(\ref{diffEqN}) will yield the same values of central moments and cumulants from second order onwards. Therefore, the PDFs obtained from any of these two can be used conveniently. However, for consistency, PDFs obtained from Eq.(\ref{diffEqN}) will be used for this study.
\par
To determine the PDF of an experimental frequency data, the first step is to calculate, $\mu$, $m_{2}, m_{3}, m_{4}$ or first four cumulants. Once the parameters are estimated using Eq(\ref{param5})-(\ref{param9}), the solution of Eq.(\ref{diffEqN}) will yield the PDF. To be noted that the PCM is applicable only when $\beta_{2} - \beta_{1} -1 > 0$. There are basically seven types of Pearson family of curves exist. Deciding the type of curves depends on the types of roots of the quadratic in the denominator of the differential equations given in Eq.(\ref{diffEq}) and Eq.(\ref{diffEqN}). Some rigorous examples of deriving the PDF from frequency data using PCM can be found in Ref.\cite{Pearson343,Pearson443,statbook}. These families of curves are related to several other known distributions. For example, Beta, Wigner semicircle, Normal, Student's t, Levy, Fisher-Snedecor, Cauchy distributions are special cases of type 1, 2, 3, 4, 5, 6 and 7 Pearson family of curves, respectively. Nevertheless, the derivation of PDFs using PCM can be done in a straightforward way using modern mathematical tools, like $Mathematica$ \cite{mathematica}, which is used for this work.
\par
PCM is a very powerful tool to derive the PDF from a set of frequency data. It also eliminates the arbitrariness of using different PDF to the same frequency data. PCM is widely used in Economics, Bio-Science, Engineering for fitting frequency data and to approximate the PDF. A probability distribution can't describe uniquely by the first four cumulants (moments). PCM can be used to model other probability distributions other than its family of curves. It can exactly reproduce the first four cumulants. However, it does not guarantee to reproduce the higher-order cumulants for $n > 4$. In reality, the knowledge about the origin and the exact probabilistic nature of the data is not known as {\it 'a priori'}. So far the detector acceptance and efficiency corrected net-proton multiplicity distribution is not obtained. Meanwhile, none of the statistical models (Skellam and Negative Binomial distributions) are successful in describing the experimental results of the individual particle and net-proton multiplicity distributions. Hence, PCM provides an alternative scope for modeling the multiplicity distributions and estimating the higher order cumulants for the baseline study.

\section{Results and discussion}
The analysis is carried out by using the beam energy dependence results of first four cumulants of proton (p), anti-proton($\bar{\textrm{p}}$) and net-proton ($\Delta \textrm{p}$) multiplicity fluctuations measured at midrapidity $|y| < 0.5$, for the transverse momentum ($p_{T}$) range $0.4 < p_{T} < 2.0$ GeV/$c$ by STAR experiment at RHIC \cite{STAR:2021iop}. In this article, the analysis is focused only on most central collision events (0-5$\%$ centrality).
\par
First, $\mu$, $m_{2}, m_{3}$, and $m_{4}$ are estimated from the detector efficiency corrected results of the first four cumulants of p, $\bar{\textrm{p}}$ and $\Delta \textrm{p}$ multiplicity distributions. The constant parameters of Eq.(\ref{diffEqN}) are estimated by invoking Eq.(\ref{param5})-(\ref{param9}). Then the functional form of the PDFs are derived using PCM.

\subsection{PDF of proton multiplicity distributions}
For Au-Au collisions at $\sqrt{s_{NN}}$ = 7.7, the PDF $f(x)$, of proton multiplicity distributions is found to be type 4 Pearson family of curve, which is expressed in Eq. (\ref{eq1}).
\begin{equation}
f(x) = \frac{1}{\sqrt{\pi}}\frac{e^{1197.11+195.271~tanh^{-1}( 0.217 + 0.012x)}}{(7363.63 + 36.3573x+x^{2})^{142.433}}.
\label{eq1}
\end{equation}

Here $x$ represents the proton multiplicity. Type 4 Pearson family of curves is not a standard distribution. For $\sqrt{s_{NN}}$ = 14.5 GeV,  the proton multiplicity distribution is constructed as type 6 Pearson family of curves. The PDF is a shifted and rescaled Fisher-Snedecor distribution. It can be expressed as follows.
\begin{equation}
f(x) = \frac{(a - b )^{\alpha - 1}(x - b)^{\zeta}}{B(\alpha -1, 1 - \zeta) (x - a)^{\zeta}}~,
\end{equation}
for all $x > a$. Here $\zeta = (\beta+\alpha a)/(a-b)$, with $a =-42.6642, b = -302.706, \alpha = 920.866$ and $\beta = -24651.3$.

\par
For $\sqrt{s_{NN}}$ = 11.5, 19.6, 27, 39, 54.4, 62.4 and 200 GeV, the proton multiplicity distribution are constructed as type 1 Pearson family of curve, which are shifted and rescaled (SR) Beta distribution. The Beta distribution is expressed as follows.
\begin{equation}
f(x) = \frac{ x^{\alpha - 1}(1-x)^{\beta-1}}{B(\alpha, \beta)} ; x \in (0,1)~.
\end{equation}
Here the normalisation constant $B(\alpha, \beta)$ is Beta function, which is given as,
\begin{equation}
B(\alpha, \beta) = \frac{\Gamma(\alpha) \Gamma(\beta)}{\Gamma(\alpha+\beta)}~,
\end{equation}
where $\alpha$ and $\beta$ are shape parameters. The Beta distribution is the conjugate prior probability distribution for the Bernoulli, (Negative)Binomial Distributions (NBD/BD) and Geometric distributions. The SR Beta distribution is obtained by performing a transformation, $x \rightarrow a + (b-a)x$, which can be written as,
\begin{equation}
f(x) = \frac{ (x-a)^{\alpha - 1}(b-x)^{\beta-1}}{B(\alpha, \beta) (b-a)^{\alpha+\beta-1}}; x \in [a,b]~.
\label{betaP}
\end{equation}
In Eq.(\ref{betaP}), $a$ and $b$ are constant parameters. The values of these constant parameters obtained for $\sqrt{s_{NN}}$ = 11.5, 19.6, 27, 39, 54.4, 62.4 and 200 GeV are given in Table \ref{tabp}.

\begin{table}[ht]
\centering
\caption{ Values of the different constant parameters of the shifted and rescaled Beta functions constructed for proton multiplicity distributions for different collision energies.} 
\label{tabp}
 \begin{tabular}{c | c | c | c |c }
  \hline
$\sqrt{s_{NN}}$ & $a$ & $b$ & $\alpha$ & $\beta$ \\
\hline
11.5 & -20.179 & 236.093 & 73.615 & 293.446 \\
 19.6 & -10.128 & 121.403  & 39.351 & 110.805  \\ 
 27 & -10.504 &  120.919 &  39.328 &  122.724  \\
 39 &  -10.944  & 129.012  &  39.763 &  145.223 \\
 54 & -16.033 & 235.078 &  59.736  & 372.315 \\
 62.4 &  -16.746 &  374.089 & 65.315  & 667.012 \\
 200 &  -11.02 & 140.058 & 39.69 &  161.193 \\
 \hline
\end{tabular}
\end{table}

\begin{figure}
\includegraphics[width=\linewidth]{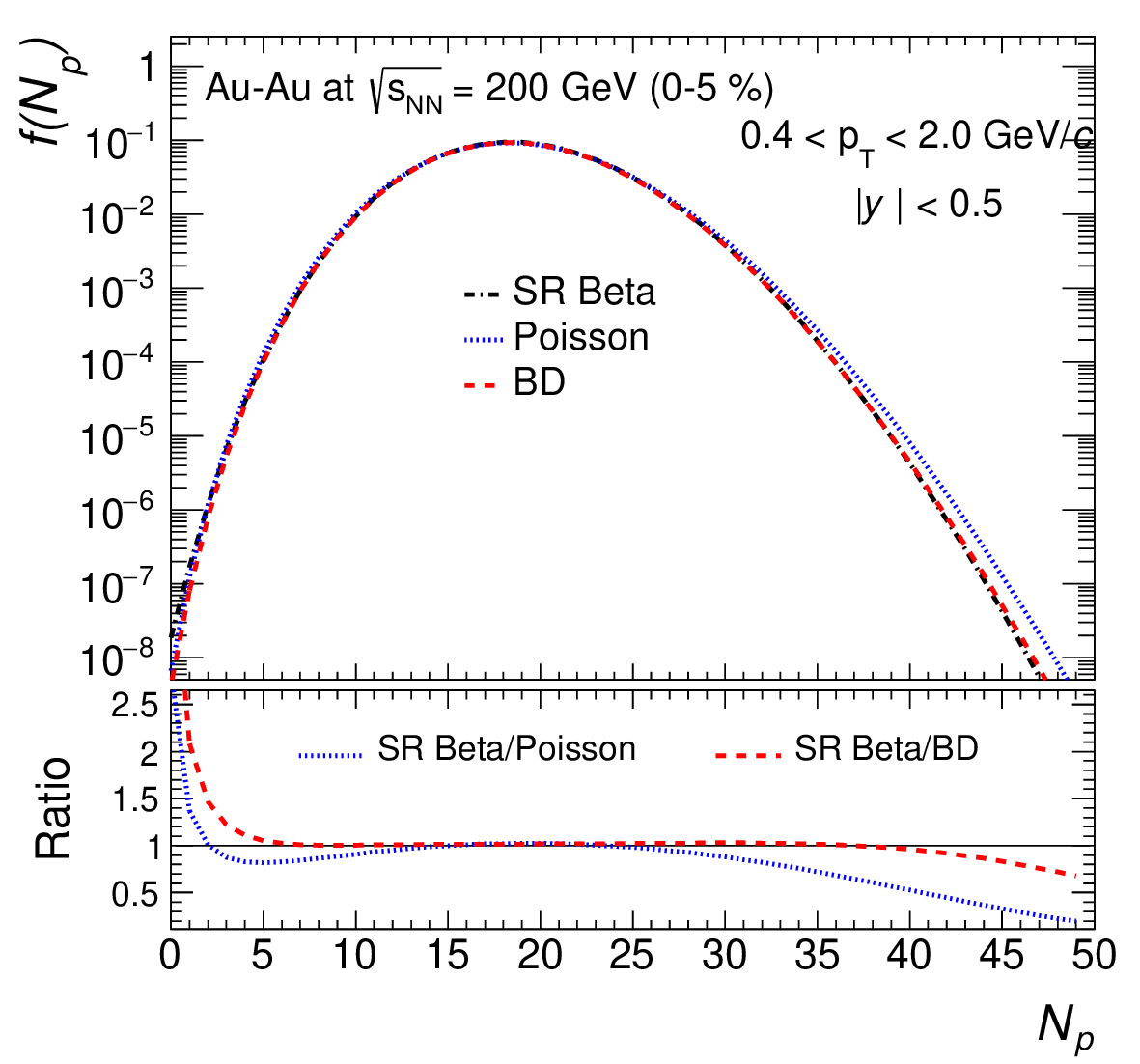}
\caption{(Color online) The constructed probability distribution of proton multiplicity from PCM using its experimental results of first four cumulants in the most central Au-Au collisions at $\sqrt{s_{NN}}$ = 200 GeV. The obtained PDF is represented by shifted and rescaled (SR) Beta distributions illustrated by dashed-dotted line. The distribution is compared with Poisson and Binomial distributions (BD), which are illustrated by dotted and dashed lines, respectively. The ratio of SR Beta distribution with respect to Poisson and BD distributions are shown in the lower panel by the dotted and dashed lines, respectively.}
\label{fig2}
\end{figure}

For illustration purpose, the probability distribution of proton multiplicity is plotted only for most central Au-Au collisions at $\sqrt{s_{NN}}$ = 200 GeV, which is shown in Figure \ref{fig2}. The distribution is also compared with Poisson and BD. A Poisson distribution is a single parameter function of $\mu$, which is expressed as follows.
\begin{equation}
f(x) = e^{-\mu}\frac{\mu^{x}}{x!}
\end{equation}
For Poisson distribution, the variance, $C_{2}$ is equal to $\mu$. The probability function of a BD distribution of a random variable $x$ is expressed in terms of the number of trials ($n$) and probability of success ($p$) as follows.
\begin{equation}
f(x) = {n\ \choose k\ } p^{x}(1-p)^{n-x}
\label{bd}
\end{equation}
For BD, $\mu = np$ and $C_{2} = np(1-p)$. The ratios of SR Beta distributions with respect to Poisson and BD are shown in the lower panels of Figure \ref{fig2}. From the ratio, it can be seen that the SR Beta distribution obtained for proton multiplicity is wider than the Poisson distribution at very small multiplicity ($N_p <$ 10). In the mid-range of multiplicity all the distribution seem indistinguishable. However, for large multiplicity $N_p >$ 27, narrowing of the distribution with respect to Poisson distribution occurs. From the ratio it is clear that the SR Beta distribution for proton is closer to BD than the Poisson distributions.

\subsection{PDF of anti-proton multiplicity distributions}
The PDF of anti-proton multiplicity distributions are constructed using PCM and found to be shifted and rescaled Beta distribution for $\sqrt{s_{NN}}$ = 7.7 - 200 GeV as given in Eq.(\ref{betaP}). The numerical values of the constant parameters of the SR Beta distribution for anti-proton multiplicity distributions for different collision energy are given in Table \ref{tabpbar}.
\begin{table}[htb]
\centering
\caption{ Numerical values of the different constant parameter of the shifted and rescaled Beta functions constructed for anti-proton multiplicity distributions for different collision energies.} 
\label{tabpbar}
 \begin{tabular}{c | c | c | c |c }
  \hline
$\sqrt{s_{NN}}$ & $a$ & $b$ & $\alpha$ & $\beta$ \\
\hline
7.7 &  -0.13 & 3.595 & 0.43 & 3.353 \\
11.5 & -0.531 & 9.465 & 1.854 &  10.552 \\
14.5 & -0.816 & 13.248 & 2.864 & 14.181 \\
19.6 & -1.443 & 21.389 & 5.137 & 24.078 \\ 
27 & -2.138 & 29.546 & 7.646 & 32.338 \\
39 & -3.263 & 44.796  & 11.714 & 51.288 \\
54.4 & -4.597 & 61.003 & 16.481 & 72.318 \\
62.4 & -5.191 & 73.611 & 18.826 & 93.556 \\
200 & -7.926 & 100.635 & 28.494 & 110.911 \\
 \hline
\end{tabular}
\end{table}

\begin{figure}
\includegraphics[width=\linewidth]{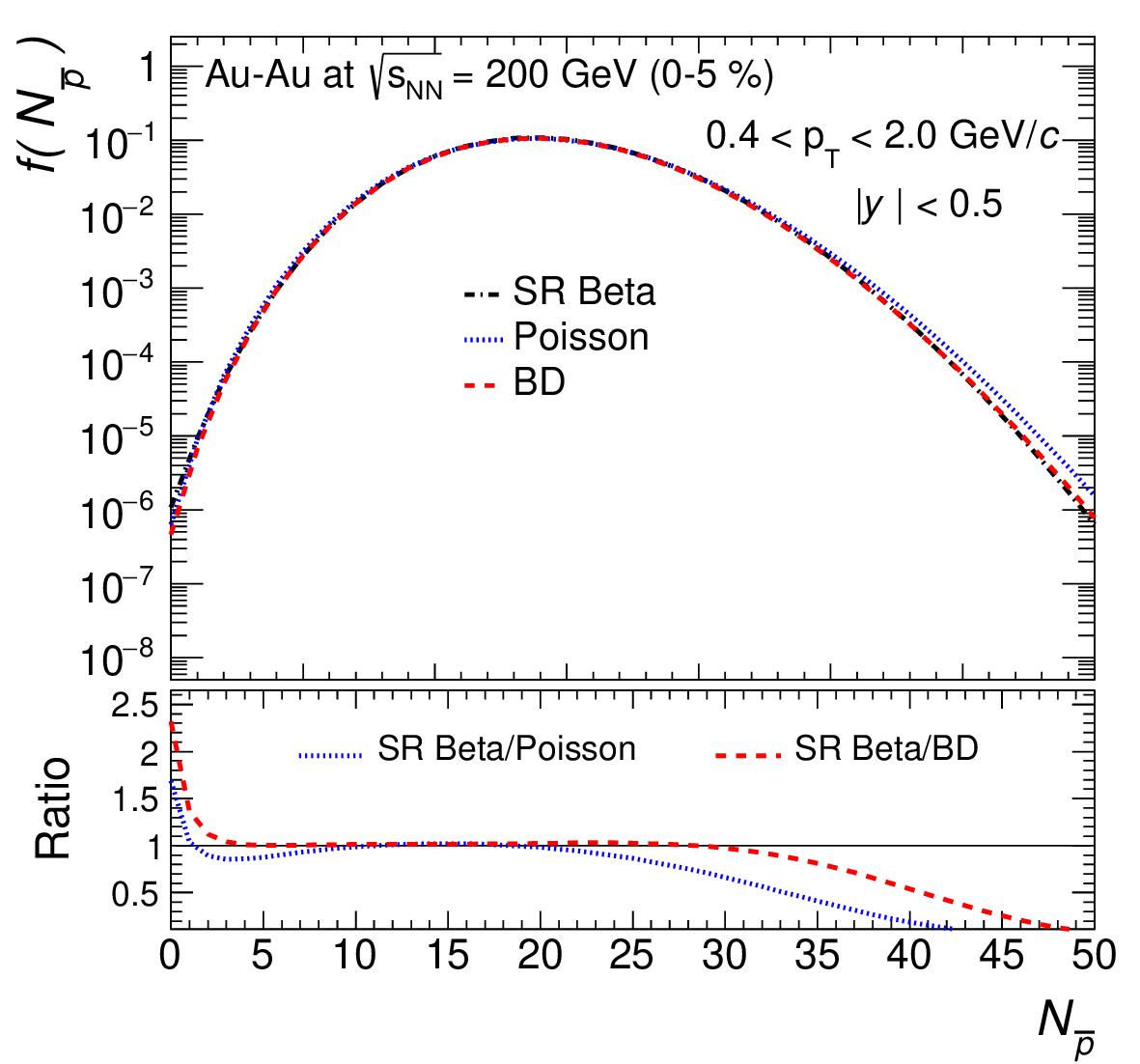}
\caption{(Color online) The constructed probability distribution of anti-proton multiplicity from the PCM using its experimental results of the first four cumulants in the most central Au-Au collisions at $\sqrt{s_{NN}}$ = 200 GeV. The obtained PDF is represented by shifted and rescaled (SR) Beta distributions, which is illustrated by dashed-dotted line. The distribution is compared with Poisson and Binomial distributions (BD), which are illustrated by the dotted and dashed lines, respectively. The ratio of SR Beta distribution with respect to Poisson and BD distributions are shown in the lower panel by the dotted and dashed line, respectively.}
\label{fig3}
\end{figure}

The probability distribution (SR Beta distribution) of anti-proton obtained from PCM for Au-Au collisions at  $\sqrt{s_{NN}}$ = 200 GeV is compared with Poisson and BD distribution, which are illustrated in the upper panel of Figure \ref{fig3}. The difference of SR Beta distribution with respect to Poisson and BD is estimated by taking the ratios and shown in the lower panel of Figure \ref{fig3}. From the ratio, it is observed that the constructed anti-proton distribution has a wider tail than the Poisson and BD at lower multiplicity range ($N_{\bar{p}} < $ 7). With increasing the multiplicity all the distributions almost merge. However, at large value of anti-proton multiplicity ($N_{\bar{p}} > $ 23), it gets narrower with respect to the Poisson distribution. At $\sqrt{s_{NN}}$ = 200 GeV, like the proton multiplicity distribution, the SR Beta distribution for anti-proton is closer to BD than the Poisson distributions.

\subsection{PDF of net-proton proton multiplicity distributions}
The PDF of net-proton multiplicity distributions in Au-Au collisions at $\sqrt{s_{NN}}$ = 7.7, 14.5, 39, 54.4, 62.4 and 200 GeV are constructed as type 4 Pearson family of curve, which are given in Eq.(\ref{pdf7hpt}), (\ref{pdf14hpt}), (\ref{pdf39hpt}),  (\ref{pdf54hpt}), (\ref{pdf62hpt}) and (\ref{pdf200hpt}), respectively.
\begin{equation}
f(x) = \frac{1}{\sqrt{\pi}}\frac{e^{ 1186.02 + 180.332~tanh^{-1}( 0.184 + 0.012 x)}}{(7445.21 + 31.301x + x^{2})^{139.926}}.
\label{pdf7hpt}
\end{equation}
\begin{equation}
f(x) = \frac{1}{\sqrt{\pi}}\frac{e^{1128.99 + 246.242~tanh^{-1}( 0.462 + 0.015 x)}}{(5342.11 + 61.339x+x^{2})^{146.017}}.
\label{pdf14hpt}
\end{equation}

\begin{equation}
f(x) = \frac{1}{\sqrt{\pi}}\frac{e^{1304.06 + 135.835~tanh^{-1}( 0.266+ 0.0132x)}}{(6160.13 + 40.3223x+x^{2})^{154.198}}.
\label{pdf39hpt}
\end{equation}

\begin{equation}
f(x) = \frac{1}{\sqrt{\pi}}\frac{e^{1393.37 + 105.995~tanh^{-1}(0.198 + 0.012x)}}{( 7044.98 + 32.6299x+x^{2})^{160.123}}.
\label{pdf54hpt}
\end{equation}

\begin{equation}
f(x) = \frac{1}{\sqrt{\pi}}\frac{e^{971.675 + 58.9903~tanh^{-1}(0.116+ 0.014 x)}}{( 5125.62 + 16.58x+x^{2})^{115.036}}.
\label{pdf62hpt}
\end{equation}

\begin{equation}
f(x) = \frac{1}{\sqrt{\pi}}\frac{e^{920.661 + 16.782~tanh^{-1}(0.022 + 0.012x)}}{(6251.8 + 3.544x+x^{2})^{105.648}}.
\label{pdf200hpt}
\end{equation}

In these equations, $x$ represents net-proton multiplicity. The PDF of net-proton multiplicity distributions for Au-Au collisions at $\sqrt{s_{NN}}$ =  11.5, 19.6 and 27 GeV are obtained as SR Beta distributions. The constant parameters of the SR Beta functions for the two energies are listed in Table \ref{tabNetp} .

\begin{table}[ht]
\centering
\caption{ Numerical values of the different constant parameter of the shifted and rescaled Beta functions constructed for net-multiplicity distributions for Au-Au collisions at $\sqrt{s_{NN}}$ = 11.5, 19.6 and 27 GeV.} 
\label{tabNetp}
 \begin{tabular}{c | c | c | c |c }
  \hline
$\sqrt{s_{NN}}$ & $a$ & $b$ & $\alpha$ & $\beta$ \\
\hline
11.5 & -31.71 & 398.519 & 112.311 & 667.495 \\
19.6 & -27.913 & 168.932 & 76.134 & 225.515 \\ 
27 & -40.7822 & 219.304 & 115.847 & 401.39 \\
 \hline
\end{tabular}
\end{table}

\begin{figure}
\includegraphics[width=\linewidth]{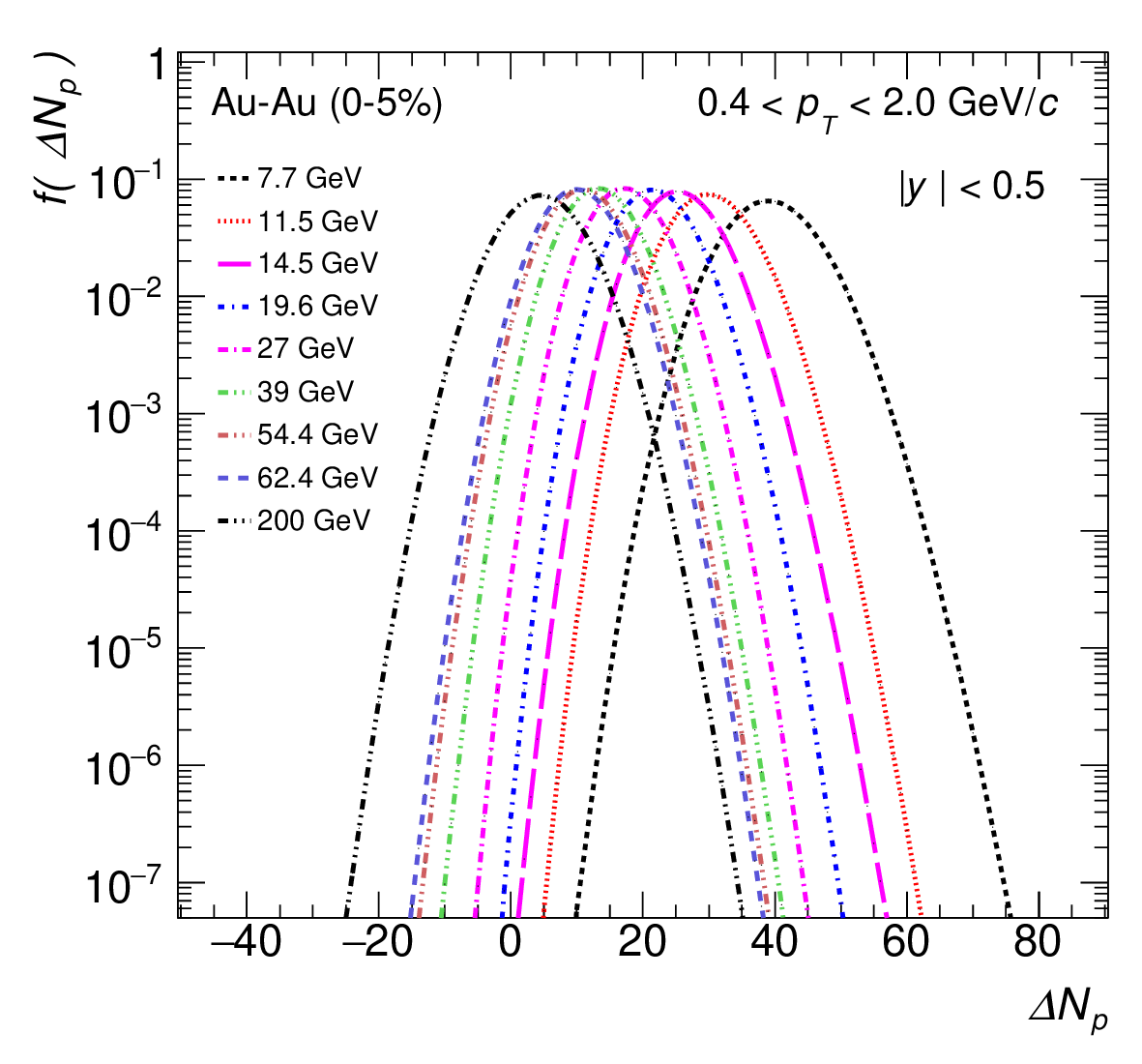}
\caption{(Color online) The constructed probability distribution of net-proton multiplicity from the PCM using its experimental results of first four cumulants in the most central Au-Au collisions at $\sqrt{s_{NN}}$ = 7.7 - 200 GeV.}
\label{fig4}
\end{figure}

The probability distributions constructed for the net-proton multiplicity distributions in the most central Au-Au collisions at $\sqrt{s_{NN}}$ = 7.7 - 200 GeV are illustrated in Figure \ref{fig4}. It can be seen that the net-proton distribution becomes symmetric around zero while going from 7.7 GeV to 200 GeV.

\subsection{Test for the criticality of the net-proton multiplicity distributions}
The probability distribution of a conserved charge is the fundamental of its cumulants. So the signature of criticality near the QCD phase boundary will be reflected both in its higher order cumulants and in the probability distributions. To demonstrate the Z(2) and O(4) criticality, first the net-quark probability distributions and later the net-baryon probability distributions at zero and non-zero $\mu_{B}$ are investigated \cite{Morita:2013tu,Morita:2012kt,Morita:2014fda}. By comparing the probability distributions with non-singular distribution (Skellam function), it is shown that the deviation with respect to the Skellam function can be used to study the criticality. Using Landau theory of phase transition it is shown that for O(4) universality the probability distribution is characterised by narrower for intermediate values of $|\Delta N_{p}|$ and a wider tail at large value of $|\Delta N_{p}|$ with respect to the Skellam distribution \cite{Morita:2012kt}. The study done in Ref. \cite{Morita:2013tu} using the Functional Renormalization Group approach to the Quak-Meson (QM) model shows that the narrowing of the probability distribution relative to the Skellam function can be regarded as a signature of O(4) criticality near the chiral crossover transition. As a consequence, the sixth order cumulant of net-baryon will be negative. Furthermore, it shows that the narrowing is due to the effect Fermi statistics \footnote{The Skellam distribution used in Ref. \cite{Morita:2012kt} is different than the one used in Ref. \cite{Morita:2014fda}}.

Higher order cumulants are sensitive to the tail of the probability distribution. Hence, the knowledge of the distributions for a sufficiently large value of $|\Delta N_{p}|$ is required for the exact determination of $C_{6}$ and its characteristics properties. The value of $|\Delta N_{p}|$ increases with increasing the order of cumulant. In Ref. \cite{Morita:2014fda}, the probability distributions ($P(N)^{FRG}$) of net-baryon is obtained from the FRG approach within the QM model at zero and non-zero $\mu_{B}$. Then the ratio of the $P(N)^{FRG}$ and the Skellam distribution ($P(N)^{S}$) is computed as a function of shifted mean ($\delta N_{p} = \Delta N_{p} - \mu$) normalised to $N_{6}$. Here, $N_{6}$ is the maximum value of $|\Delta N_{p}|$ where the $C_{6}$ is exactly reproduced from the Skellam distribution to its exact value \cite{Morita:2013tu}. In this case, the Skellam distribution has the same mean and variance as the distribution it is compared with. Then the efficiency uncorrected net-proton probability distributions reported by the STAR experiment in the kinematic range $0.4 < p_{T} < 0.8$ GeV/$c$ and $|y| < 0.5$, are used to investigate any possible O(4) criticality. The study shows some qualitatively similar structure as predicted in the model \cite{Adamczyk:2013dal,Morita:2014fda}.

\begin{figure}
\includegraphics[width=\linewidth]{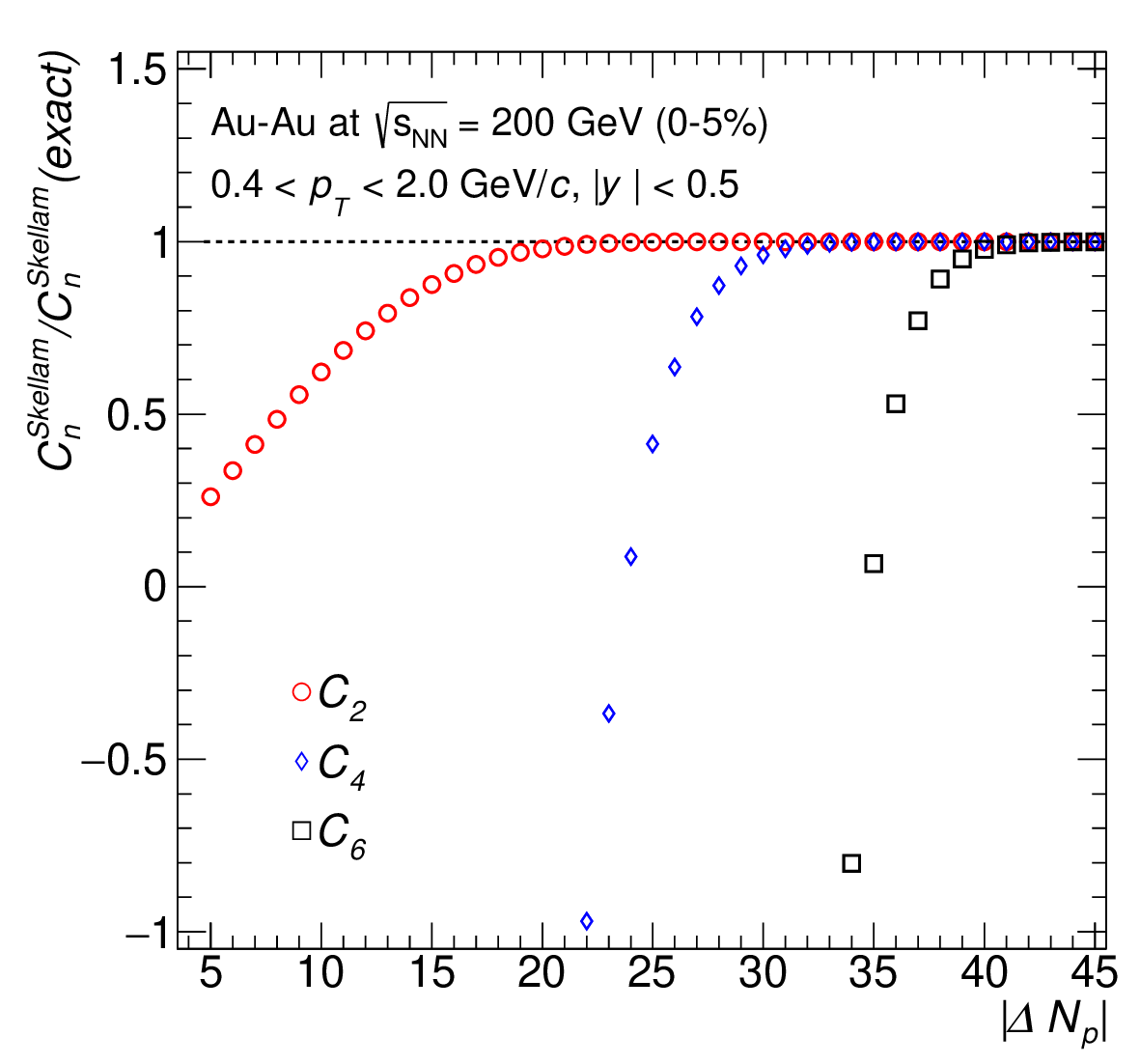}
\caption{(Color online) The ratios of $2^{nd}, 4^{th}$ and $6^{th}$ order cumulants of the Skellam distribution to their exact values as a function of $|\Delta N_{p}|$ for the most central Au-Au collisions at $\sqrt{s_{NN}}$ = 200 GeV.}
\label{Nmax200}
\end{figure}

Here, an attempt is made to carry out similar exercise to investigate any possible O(4) criticality associated with the constructed probability distributions ($f(\Delta N_{p})^{PCM})$ for the efficiency corrected cumulants results using PCM in most central Au-Au collisions at $\sqrt{s_{NN}}$ = 14.5, 19.6, 27, 39, 54.4, 62.4 and 200 GeV in the kinematic range $0.4 < p_{T} < 2.0$ GeV/$c$ and $|y| < 0.5$. The functional form of the constructed probability distributions are already given in the previous section. First, the maximum value ($N_{max}$) for $2^{nd}, 4^{th}$ and $6^{th}$ order of cumulants for the Skellam distributions for different collision energies are estimated. Here the Skellam distributions are constructed with same $C_1$ and $C_2$ value as the net-proton data. For a given order of cumulant, the $N_{max}$ is determined from the value of $|\Delta N_{p}|$ at which the cumulant value converges with its exact values. The convergence is checked by taking the ratio of cumulant estimated from the distribution to its expected value. For illustration purpose the ratio at different value of $|\Delta N_{p}|$ for Au-Au collisions at $\sqrt{s_{NN}}$ = 200 GeV is shown in Figure \ref{Nmax200}. The open circle, open diamond and open square represent the ratios of the $2^{nd}, 4^{th}$ and $6^{th}$ order cumulants, respectively, for different value of $|\Delta N_{p}|$. It is clear from Figure \ref{Nmax200} that the $N_{max}$ value increases with increasing the order of cumulants. Moreover, it is observed that at relatively smaller range of $|\Delta N_{p}|$ the $C_4$ and $C_6$ can have negative sign \footnote{In fact with increasing the $|\Delta N_{p}|$, $C_6$ oscillates from very high positive value to negative value before converging with its exact value, which is not visible in the current scale of Figure \ref{Nmax200}}. A recent MC study with Skellam distribution reported negative value of $C_5$ and $C_6$ \cite{Pandav:2018bdx}. From this study it is evident that their observed negative value is purely due to insufficient event statistics, which fails to explore the required range of $\Delta N_{p}$.
\par
\begin{figure}
\includegraphics[width=\linewidth]{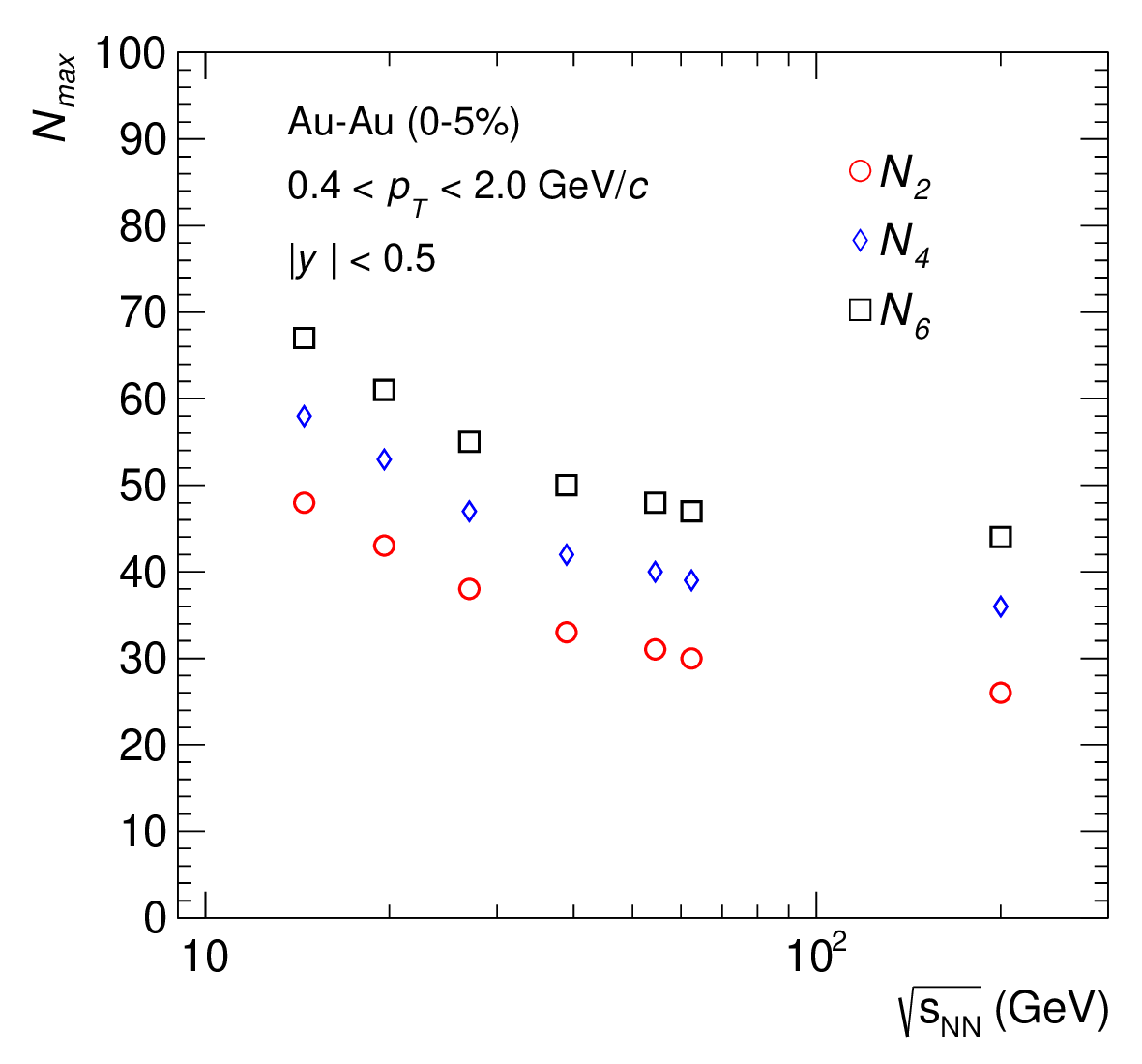}
\caption{(Color online) The $N_{max}$ value for $C_{2}$, $C_{4}$ and $C_{6}$ in most central Au-Au collision at $\sqrt{s_{NN}}$ = 14.5. 19.6, 27, 39, 54.4, 62.4 and 200 GeV. $N_{max}$ corresponds to the maximum value of $|\Delta N_{p}|$ at which the $C_n$ reproduced from the Skellam distribution is exactly same as the expected one.}
\label{Nmax}
\end{figure}

In a similar way, the $N_{max}$ values corresponding to $C_2, C_4$ and $C_6$ for other collision energies are also estimated. The  $N_{max}$ as a function of collision energy are shown in Figure \ref{Nmax}. In Figure \ref{Nmax}, the $N_{2}$, $N_{4}$ and $N_{6}$ values correspond to the maximum value of $|\Delta N_{p}|$ at which the $C_{2}$, $C_{4}$ and $C_{6}$, respectively, reproduced from the Skellam distribution as the expected ones. The numerical values of  $N_{2}$, $N_{4}$ and $N_{6}$  for different energies are also given in Table \ref{tab4}. From Figure \ref{Nmax} and Table \ref{tab4} it is clear that for a given order of cumulant, with increasing collision energy the $N_{max}$ decreases. The decrease is faster at the lower energy range. Furthermore, it is observed that for given collision energy the $N_{max}$ shows an increase by almost a constant factor with increasing the order of cumulants. A similar observation is made in Ref. \cite{Morita:2013tu}. This implies that more event statistics are required to reveal the tail and in particular the higher order cumulants of the distribution.

\begin{table}[ht]
\centering
\caption{ Values of $N_{2}, N_{4}$ and $N_{6}$ for most central Au-Au collisions at $\sqrt{s_{NN}}$ = 14.5. 19.6, 27, 39, 54.4, 62.4 and 200 GeV. } 
\label{tab4}
 \begin{tabular}{c | c |c |c }
  \hline
$\sqrt{s_{NN}}$ & ~~$N_{2}$~~& ~~$N_{4}$~~ & ~~$N_{6}$~~\\
\hline
14.5 & 48 & 58 & 67 \\
19.6 &  43 &  53 & 61 \\
27 &  38 & 47 & 55 \\
39 &  33 & 42 & 50 \\
54.4 & 31 & 40 & 48 \\
62.4 & 30 & 39 & 47 \\
200 &  26 & 36 & 44 \\
 \hline
\end{tabular}
\end{table}

\begin{figure}
\includegraphics[width=\linewidth]{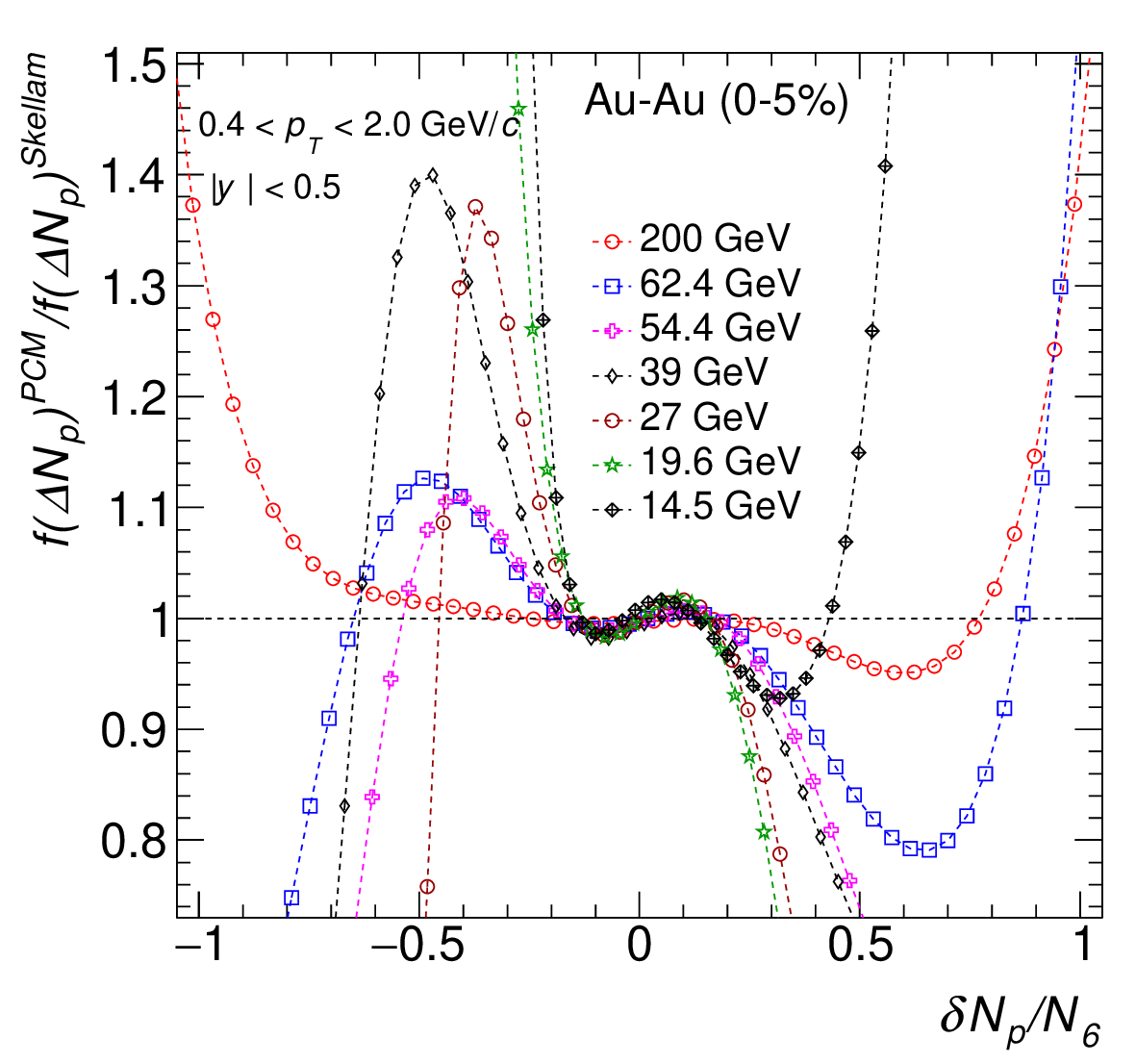}
\caption{(Color online) Ratio of the probability distribution obtained from PCM ($f(\Delta N_{p})^{PCM}$) to the Skellam distribution ($P(N)^{S}$) for most central Au-Au collisions at $\sqrt{s_{NN}}$ = 14.5, 19.6, 27, 39, 54.4, 62.4 and 200 GeV. The Skellam distribution has the same $\mu$ and $C_{2}$ as of the $f(\Delta N_{p})^{PCM}$ for a given energy. The ratio is shown for different value of $\delta N_{p}/N_{6}$.}
\label{figNormRatio}
\end{figure}

In the second step, the constructed probability distributions are compared with the Skellam distribution to find out the signature of O(4) criticality. The ratio $f(\Delta N_{p})^{PCM}/P(N)^{S}$ is estimated for different collision energies and illustrated as a function of $\delta N_{p}/N_{6}$ in Figure \ref{figNormRatio}. For Au-Au collisions at $\sqrt{s_{NN}}$ = 14.5 GeV, the ratio $f(\Delta N_{p})^{PCM}/P(N)^{S}$ shows an oscillatory behaviour at very small value of $|\delta N_{p}/N_{6}|$ value, and then increases very rapidly with increasing $|\delta N_{p}/N_{6}| < $ 0.5. For Au-Au collisions at $\sqrt{s_{NN}}$ = 19.6, 27, 39,  54.4 and 62.4 GeV, the ratio $f(\Delta N_{p})^{PCM}/P(N)^{S}$ shows a small increase for very small positive value of $\delta N_{p}/N_{6}$ value. Then it shows a sharp drop for $\delta N_{p}/N_{6} <$ 0.5. This trend is opposite for the negative value of $\delta N_{p}/N_{6}$, which shows the asymmetry of the probability distributions. The drop in the ratio occurs relatively a smaller value of $\delta N_{p}/N_{6}$ with decreasing the energy. This observation is inline with the model study done for the probability distribution constructed based on the FRG approach and with the efficiency uncorrected data \cite{Morita:2014fda}. This early drop in the ratio and asymmetry of the distribution implies that the criticality will appear at a smaller value of $\delta N_{p}$, and hence in the lower order of cumulants.
\par
For Au-Au collisions at $\sqrt{s_{NN}}$ =  200 GeV, the distributions show very small oscillatory behaviour for small $\delta N_{p}/N_{6}$ value. But with increasing the $\delta N_{p}/N_{6}$ the distribution gets narrower up to 0.5.  After that, it again broadens for both values of $\delta N_{p}$, which is reflected in the sharp increase of the ratio. The observed feature for this energy is qualitatively similar to the feature seen in the model study based on the Landau theory of phase transition \cite{Morita:2012kt}. However, this feature does not agree with the model study based on the FRG approach \cite{Morita:2014fda}. Interestingly, some of signatures of the O(4) criticality, such as narrowing of the probability distribution for $\delta N_{p}/N_{6} <$ 0.5 is observed. Hence, the possibility of remnants of the chiral phase transition in the constructed probability distributions can not be ruled out. To be noted here that the cumulant results used in the PDF constructions are not corrected for volume fluctuations and the effect of global baryon number conservation. Therefore, more insight can be gained after understanding these effect. 
\par
It is also suggested that the narrowing of the distribution due to O(4) criticality results in a negative value of sixth-order cumulant. So for a further test of O(4) criticality, the $C_{6}$ and its ratio $C_{6}/C_{2}$ for different collision energies are estimated from the PDFs, which is discussed in the following section.

\subsection{Estimating higher-order cumulants}
\begin{figure}
\includegraphics[width=\linewidth]{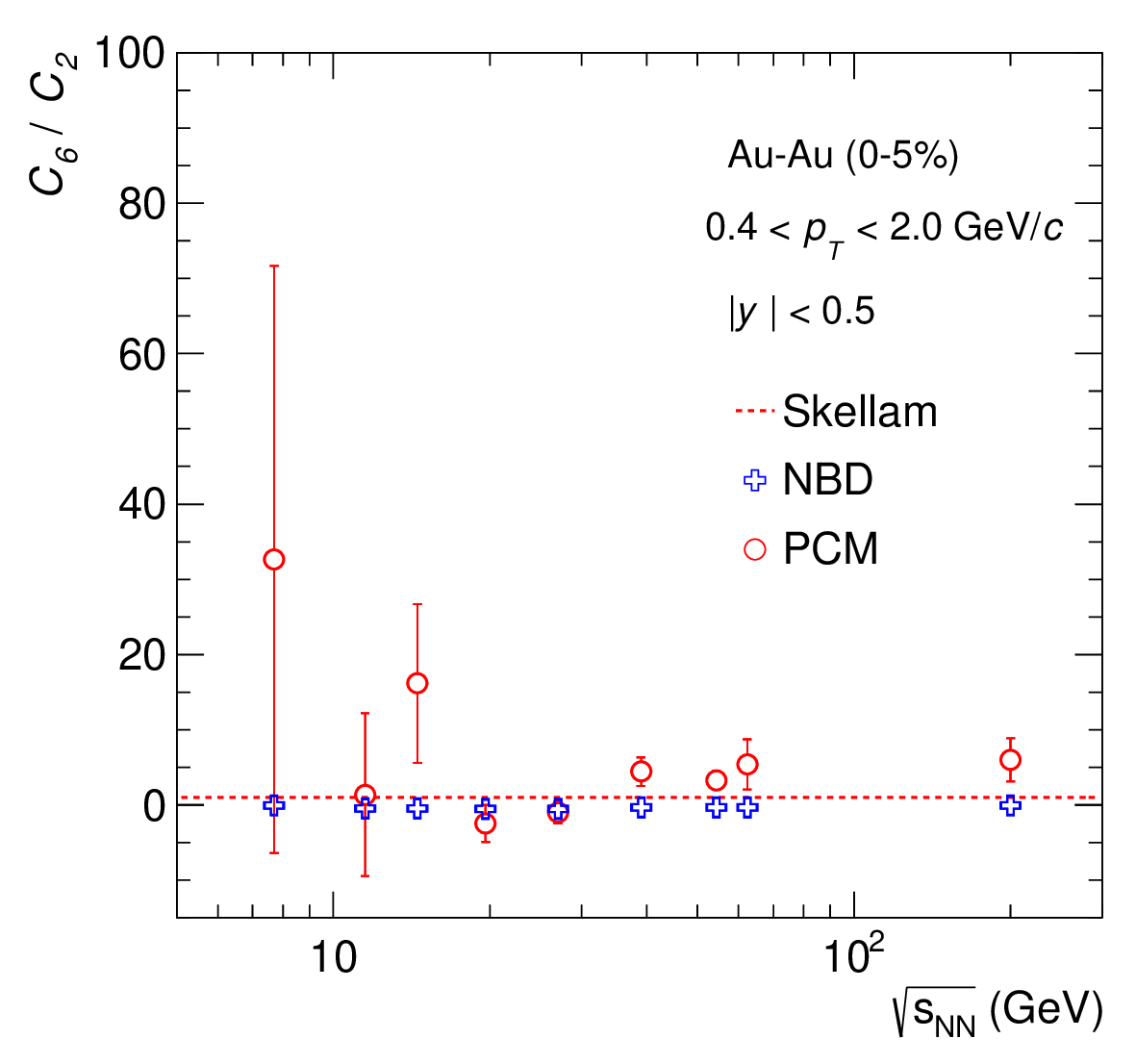}
\caption{(Color online) Energy dependence of $C_{6}/C_{2}$ of net-proton multiplicity distribution estimated from PCM for most central collisions events in $0.4 < p_{T} < 2.$ GeV/$c$ compared with Skellam and NBD expectations. The PCM  results are illustrated by the open circles. The Skellam expectation is represented by the dotted line and the open cross marker represents the NBD expectation.}
\label{figc6c2}
\end{figure}

From the obtained PDFs of net-proton multiplicity distributions the ratio, $C_6/C_2$, are calculated analytically for $\sqrt{s_{NN}}$ = 7.7 - 200 GeV. As the experimental results are associated with statistical uncertainties, their effects on the estimated $C_6/C_2$ values need to be evaluated. It should be noted here that the propagation of uncertainties from the first four cumulants to the analytically estimated $C_6/C_2$ value is rather nontrivial due to their complex relationship with the constant parameters of Eq.(\ref{param5})-(\ref{param9}), and the constructed PDF in PCM.  Therefore, we consider the following sampling method. The value of $C_{1}$, $C_{2}$, $C_{3}$, and $C_{4}$ are sampled (5000 times) randomly by varying their values within the 3$\sigma_{c}$ of their mean values, assuming independent Gaussian distributions. Here $\sigma_{c}$ refers to the statistical uncertainty of a given cumulant at a given energy. Then the constant parameters of Eq.(\ref{param5})-(\ref{param9}) and hence the analytic value of  $C_6/C_2$ are estimated for each set of $C_{1}$, $C_{2}$, $C_{3}$, and $C_{4}$ values. Finally, $C_{6}/C_{2}$ and the uncertainty on it is estimated by calculating the mean and standard deviations of all the samples, respectively. The $C_6/C_2$ result for $\sqrt{s_{NN}}$ = 7.7 - 200 GeV is shown in Figure \ref{figc6c2}. It can be observed from Figure \ref{figc6c2} that the $C_6/C_2$ result for $\sqrt{s_{NN}}$ = 7.7 GeV has the largest uncertainty, which is due to the largest statistical uncertainties associated with the input cumulants ($C_{1}$, $C_{2}$, $C_{3}$, and $C_{4}$), among all the energies. However, it is difficult to draw any direct proportionality relation between the estimated uncertainties and the uncertainties of the lower order cumulants due to the complexity mentioned above.

Furthermore, the baseline estimation is done for the BES data from the Skellam and NBD/BD distributions. When the proton and anti-proton probability density distributions are assumed to be two uncorrelated Poissonian distributions, the net-proton distribution becomes Skellam distribution. Skellam distribution is widely used in HRG model to estimate the cumulants of conserved charges \cite{Karsch:2010ck,Garg:2013ata,Braun-Munzinger:2014lba}. The $n^{th}$ cumulant of Skellam distribution is defined as  $C_{n} = c_{1}(p) + (-1)^n c_{1}(\bar{p})$. Here $c_{1}(p)$ and $c_{1}(\bar{p})$ are the first cumulant of proton and antiproton distributions, respectively. So the all even order cumulants are same, and hence their ratios, like $C_{4}/C_{2}$ and $C_{6}/C_{2}$ will always yield unity. It can be seen here that values of all orders of cumulants of net-proton distribution depend entirely on the first cumulants of proton and anti-proton distributions.
\par
Recently, NBD/BD has been proposed as one of the baselines for such study \cite{Tarnowsky:2012vu}. In this case, the proton and anti-proton distributions are assumed two uncorrelated NBD/BD. For a given NBD/BD, one can calculate the cumulants, like $C_{2}, C_4$ and $C_6$ analytically by knowing the value of the first and second cumulants \cite{Tarnowsky:2012vu}. For the BES data, proton mean value is greater than its variance ($\mu > \sigma^2$). This is also true for anti-proton. Hence, BD has to be used to approximate the proton and anti-proton multiplicity distributions. Under this assumption, the $C_n$ of net-proton multiplicity distributions are calculated from the cumulants of proton and anti-proton distributions as follows.
\begin{equation}
C_{n} (\Delta N_{p}) = c_{n}(p) + (-1)^n c_{n}(\bar{p}),
\label{cneq}
\end{equation}
where $c_{n}(p)$ and $c_{n}(\bar{p})$ are the $n^{th}$ order cumulants of the assumed BD for proton and anti-proton multiplicity distributions, respectively. 

The $C_{6}/C_{2}$ results of PCM compared with the baselines results estimated from Skellam and BD are illustrated in Figure \ref{figc6c2}.  The baseline result estimated from BD does not show any energy dependence and remains negative for all the energies. Interestingly, the $C_{6}/C_{2}$ result obtained from the constructed PDFs using PCM show clear energy dependence. It can be observed from Figure \ref{figc6c2} that the energy dependence of $C_{6}/C_{2}$ show a qualitatively similar trend to the energy dependence behavior of $C_{4}/C_{2}$ reported in Ref.\cite{STAR:2020tga,STAR:2021iop}. The BES result of $C_{6}/C_{2}$ deviates from Skellam expectations with oscillating in nature. For the energy range $\sqrt{s_{NN}} \geq $ 39 GeV, the value of $C_{6}/C_{2}$ does not show any significant change. In this energy range, the $C_{6}/C_{2}$ deviates from Skellam similar to the $C_{4}/C_{2}$ data. But in this case, the values are always greater than the Skellam expectations. For Au-Au collisions at $\sqrt{s_{NN}}$ = 19.6 and 27 GeV, the $C_{6}/C_{2}$ becomes negative, and particularly $\sqrt{s_{NN}}$ = 27 GeV result is almost same as the BD expectations. To be reminded here that for these two energies the probability distributions are constructed as SR Beta distributions, which is a conjugate prior probability distribution for the BD. So their cumulants results are close to each other. 
\par
Going from 11.5 GeV to 7.7 GeV, the $C_{6}/C_{2}$ values shows a rapid rise and deviate strongly from the Skellam and NBD expectations. At 7.7 GeV, a very large value of $C_{6}/C_{2}$ is observed. A similar high value of $C_{6}/C_{2}$ for 7.7 GeV is estimated from completely different approach \cite{Bzdak:2018uhv}.
\par
Before making more conclusive remarks on the $C_{6}/C_{2}$ result obtained from PCM, it is imperative to establish the critical behavior from the first four cumulants as they are the sole input parameters of this method. Furthermore, some recent studies show that sources having non-critical fluctuations and their contribution, like baryon stopping, participant fluctuations, and the effect of baryon number conservation, have significant contributions to the measured cumulant results \cite{Thakur:2016znw,Braun-Munzinger:2016yjz}. To quantify the effects of such contribution on the estimated $C_{6}/C_{2}$ value from PCM, the following consideration is made. If there will be 5$\%$, 10$\%$ and 20$\%$ uncertainties on $C_{2}$, $C_{3}$, and $C_{4}$ data, respectively, due to such effects for Au-Au collisions at $\sqrt{s_{NN}}$ = 200 GeV, then there will be around 35$-$40$\%$ uncertainty on the $C_6/C_2$ value. Therefore, it is crucial to have a quantitative idea about the contributions from the sources of non-critical fluctuations before making any conclusive remark on the present result.
\par
Recent measurement of $C_{6}/C_{2}$ from BES-I data by the STAR experiment shows some negative value for 0$-$10$\%$ centrality in Au-Au collisions at $\sqrt{s_{NN}}$ = 200 GeV, which indicates a smooth crossover at top RHIC energy \cite{STAR:2021rls}. The upcoming RHIC BES-II run aims for studying the $C_{6}/C_{2}$ in larger acceptance with larger event statistics, which will be helpful to validate the baseline predictions of $C_{6}/C_{2}$ obtained from this model \cite{Luo:2017faz}. 

\section{Summary}
In summary, for the first time Pearson curve method is used to derive the PDFs of proton, anti-proton and net-proton multiplicity distributions using the efficiency corrected results of Au-Au collisions at $\sqrt{s_{NN}}$ = 7.7 - 200 GeV.  It is found that for Au-Au collisions at 200 GeV, the constructed PDFs of proton and anti-proton multiplicity distributions are more close to the Binomial distributions than Poisson distribution. The constructed PDFs of net-proton distributions have potential importance in the context of the quark-meson (QM) model within the functional renormalization group approach to study the $O(4)$ criticality. Before testing the O(4) criticality of net-proton distribution, the maximum values of $|\Delta N_{p}|$ for $C_{2}$, $C_{4}$ and $C_{6}$ of Skellam distributions are estimated for the BES data. It is found that the $N_{max}$ value increases with increasing the order of cumulants. Moreover, it is shown that the $C_4$ and $C_6$ values obtained from a Skellam distribution can have a negative sign at a smaller range of $|\Delta N_{p}|$. The obtained $N_{max}$ value will provide a guideline for MC simulation involving Skellam distribution. The test for O(4) criticality is done by computing the ratio of the constructed PDF to the Skellam distributions for different values of $\delta N_{p}/N_{6}$. For 200 GeV the distribution is found narrower than the Skellam distribution for $\delta N_{p}/N_{6} < 0.5$, and gets broader after that. At lower energies (19.6, 27 and 39, 54.4, 62.4 GeV) the distribution is found asymmetric and gets narrower at a very small range of $\delta N_{p}/N_{6}$. From this ratio, some of the signatures of O(4) criticality, such as narrowing of the distribution is observed. For further investigation, the $C_6/C_2$ is estimated from the constructed PDF of the BES data and compared with Skellam and NBD expectations. The beam energy dependence of $C_6/C_2$ value has a qualitatively similar trend as the $C_4/C_2$ data. The ratio deviates from the Skellam and NBD expectation. Only at 19.6 and 27 GeV, the ratio has a negative value. A very large positive value is observed for 7.7 GeV. The lower energy results ($\sqrt{s_{NN}} \leq $ 19  GeV) have large uncertainties. At present, it still remains unclear about the signature of O(4) criticality from the constructed PDFs. It needs further investigation by considering contributions from non-critical fluctuations, like volume fluctuations, baryon stopping and global baryon number conservations. A quantitative comparison between the upcoming  RHIC BES-II run data and PCM results will helpful to validate the baseline predictions. In addition to that, the derived PDFs can be used for model study and qualitative comparison with other theoretical predictions.
 
\begin{acknowledgements}
The authors thanks the STAR Collaboration for providing the preliminary data during the initial phase of this study. The author acknowledges R. Bellweid, S. Dash, K. Morita and B. K. Nandi  for their valuable suggestions during the preparation of this manuscript. This work was supported by the National Research Foundation of Korea (NRF) grant funded by the Korea government (MSIT) (No. 2018R1A5A1025563).
\end{acknowledgements}


\begin{thebibliography}{}
%
%
\bibitem{Pisarski:1983ms}R.~D.~Pisarski and F.~Wilczek,
  Phys.\ Rev.\ D {\bf 29}, 338 (1984).

\bibitem{Aoki:2006we}Y.~Aoki, G.~Endrodi, Z.~Fodor, S.~D.~Katz and K.~K.~Szabo,
  Nature {\bf 443}, 675 (2006).

\bibitem{Ejiri:2008xt}S.~Ejiri,
  Phys.\ Rev.\ D {\bf 78}, 074507 (2008).

\bibitem{Bowman:2008kc} 
  E.~S.~Bowman and J.~I.~Kapusta,
  Phys.\ Rev.\ C {\bf 79}, 015202 (2009).

\bibitem{Stephanov:2007fk}M.~A.~Stephanov,
  PoS LAT {\bf 2006}, 024 (2006).

\bibitem{Asakawa:1989bq}M.~Asakawa and K.~Yazaki,
  Nucl.\ Phys.\ A {\bf 504}, 668 (1989).

\bibitem{Hatta:2002sj} 
  Y.~Hatta and T.~Ikeda,
  Phys.\ Rev.\ D {\bf 67}, 014028 (2003).

\bibitem{Scavenius:2000qd} 
  O.~Scavenius, A.~Mocsy, I.~N.~Mishustin and D.~H.~Rischke,
  Phys.\ Rev.\ C {\bf 64}, 045202 (2001).

\bibitem{Halasz:1998qr} 
  A.~M.~Halasz, A.~D.~Jackson, R.~E.~Shrock, M.~A.~Stephanov and J.~J.~M.~Verbaarschot,
  Phys.\ Rev.\ D {\bf 58}, 096007 (1998).

\bibitem{Stephanov:1999zu}M.~A.~Stephanov, K.~Rajagopal and E.~V.~Shuryak,
 Phys.\ Rev.\ D {\bf 60}, 114028 (1999).

\bibitem{Stephanov:1998dy}M.~A.~Stephanov, K.~Rajagopal and E.~V.~Shuryak,
 Phys.\ Rev.\ Lett.\  {\bf 81}, 4816 (1998).

\bibitem{Stephanov:2008qz} 
  M.~A.~Stephanov,
  Phys.\ Rev.\ Lett.\  {\bf 102}, 032301 (2009).

\bibitem{Karsch:2010ck} 
  F.~Karsch and K.~Redlich,
  Phys.\ Lett.\ B {\bf 695}, 136 (2011).

\bibitem{Friman:2011pf} 
  B.~Friman, F.~Karsch, K.~Redlich and V.~Skokov,
  Eur.\ Phys.\ J.\ C {\bf 71}, 1694 (2011).

\bibitem{Bazavov:2012vg} 
  A.~Bazavov {\it et al.},
  Phys.\ Rev.\ Lett.\  {\bf 109}, 192302 (2012).

\bibitem{Borsanyi:2013hza} 
  S.~Borsanyi, Z.~Fodor, S.~D.~Katz, S.~Krieg, C.~Ratti and K.~K.~Szabo,
  Phys.\ Rev.\ Lett.\  {\bf 111}, 062005 (2013).

\bibitem{Borsanyi:2014ewa} 
  S.~Borsanyi, Z.~Fodor, S.~D.~Katz, S.~Krieg, C.~Ratti and K.~K.~Szabo,
  Phys.\ Rev.\ Lett.\  {\bf 113}, 052301 (2014).

\bibitem{Hatta:2003wn} 
  Y.~Hatta and M.~A.~Stephanov,
  Phys.\ Rev.\ Lett.\  {\bf 91}, 102003 (2003).

\bibitem{Adamczyk:2013dal} 
  L.~Adamczyk {\it et al.} [STAR Collaboration],
  Phys.\ Rev.\ Lett.\  {\bf 112}, 032302 (2014).

\bibitem{Luo:2015ewa} 
  X.~Luo [STAR Collaboration],
  PoS CPOD {\bf 2014}, 019 (2015).
  
  \bibitem{STAR:2020tga}
J.~Adam \textit{et al.} [STAR],
Phys. Rev. Lett. \textbf{126}, no.9, 092301 (2021).
  
  \bibitem{STAR:2021iop}
M.~Abdallah \textit{et al.} [STAR],
Phys. Rev. C \textbf{104}, no.2, 024902 (2021).

\bibitem{Bleicher:1999xi} 
  M.~Bleicher {\it et al.},
  J.\ Phys.\ G {\bf 25}, 1859 (1999).
  
  \bibitem{Bzdak:2016jxo}
A.~Bzdak, V.~Koch and V.~Skokov,
Eur. Phys. J. C \textbf{77}, no.5, 288 (2017).

\bibitem{Morita:2013tu} 
  K.~Morita, B.~Friman, K.~Redlich and V.~Skokov,
  Phys.\ Rev.\ C {\bf 88}, no. 3, 034903 (2013).

\bibitem{Morita:2012kt} 
  K.~Morita, V.~Skokov, B.~Friman and K.~Redlich,
  Eur.\ Phys.\ J.\ C {\bf 74}, 2706 (2014).

\bibitem{Morita:2014fda} 
  K.~Morita, B.~Friman and K.~Redlich,
  Phys.\ Lett.\ B {\bf 741}, 178 (2015).

\bibitem{Pearson343}Karl~Pearson,
 Philosophical Transactions of the Royal Society of London A: Mathematical, Physical and Engineering Sciences, {\bf 186}, 343--414 (1895).

\bibitem{Pearson443}Karl~Pearson,
 Philosophical Transactions of the Royal Society of London A: Mathematical, Physical and Engineering Sciences, {\bf 197}, 443--459 (1901).

\bibitem{statbook}J.~H.~Pollard, ``A Handbook of Numerical and statistical techniques,'' Cambridge University Press, ISBN 0 521 29750 (1977).

\bibitem{pearsontypes}O.~Podladchikova, B.~Lefebvre, V.~Krasnoselskikh and V.~Podladchikov,
 Nonlinear Processes in Geophysics, European Geosciences Union (EGU), {\bf 10}, 323-333 (2003).

\bibitem{mathematica}Wolfram~Research, Inc., Mathematica, Version 11.2, Champaign, IL (2017).

\bibitem{Pandav:2018bdx} 
  A.~Pandav, D.~Mallick and B.~Mohanty,
 Nucl.\ Phys.\ A {\bf 991}, 121608 (2019).

\bibitem{Garg:2013ata} 
  P.~Garg, D.~K.~Mishra, P.~K.~Netrakanti, B.~Mohanty, A.~K.~Mohanty, B.~K.~Singh and N.~Xu,
  Phys.\ Lett.\ B {\bf 726}, 691 (2013).

\bibitem{Braun-Munzinger:2014lba} 
  P.~Braun-Munzinger, A.~Kalweit, K.~Redlich and J.~Stachel,
  Phys.\ Lett.\ B {\bf 747}, 292 (2015).

\bibitem{Tarnowsky:2012vu} 
  T.~J.~Tarnowsky and G.~D.~Westfall,
  Phys.\ Lett.\ B {\bf 724}, 51 (2013).

\bibitem{Bzdak:2018uhv} 
  A.~Bzdak, V.~Koch, D.~Oliinychenko and J.~Steinheimer,
  Phys.\ Rev.\ C {\bf 98}, no. 5, 054901 (2018).
  
\bibitem{Thakur:2016znw} 
  D.~Thakur, S.~Jakhar, P.~Garg and R.~Sahoo,
  Phys.\ Rev.\ C {\bf 95}, no. 4, 044903 (2017).
   
\bibitem{Braun-Munzinger:2016yjz} 
  P.~Braun-Munzinger, A.~Rustamov and J.~Stachel,
  Nucl.\ Phys.\ A {\bf 960}, 114 (2017).
  
  \bibitem{STAR:2021rls}
M.~Abdallah \textit{et al.} [STAR],
Phys. Rev. Lett. \textbf{127}, 262301 (2021).
  
  \bibitem{Luo:2017faz} 
  X.~Luo and N.~Xu,
  Nucl.\ Sci.\ Tech.\  {\bf 28}, no. 8, 112 (2017).

\end{thebibliography}


\end{document}